\documentclass[a4paper,11pt,onecolumn,allowfontchangeintitle, accepted=2025-02-26]{quantumarticle}
\pdfoutput=1

\usepackage[numbers,sort&compress]{natbib}
\usepackage{hyperref}
\usepackage{xurl} 
\usepackage{url}
\usepackage{doi}
\hypersetup{
    breaklinks=true,
    colorlinks=true,
    linkcolor=blue,
    citecolor=blue,
    urlcolor=blue
}

\usepackage[utf8]{inputenc}
\usepackage{graphicx, float, amsmath, physics}
\usepackage{amssymb, amsfonts, latexsym, amsthm, bbm, braket}
\usepackage{xcolor}
\usepackage{subfigure}
\usepackage{enumerate} 
\usepackage{dsfont} 
\usepackage{tcolorbox}
\usepackage{ragged2e}
\usepackage{tikz}
\usepackage[normalem]{ulem}

\usepackage{hyperref}
\hypersetup{
    colorlinks = true,
    allcolors=blue}
\usepackage{cleveref}

\def\A{\mathcal{A}}
\def\B{\mathcal{B}}
\def\E{\mathcal{E}}
\def\K{\mathcal{K}}
\def\P{\mathrm{P}}
\def\a{\mathbf a}
\def\k{\mathbf k}
\def\t{\mathbf t}
\def\z{\mathbf z}
\def\r{\mathsf r}
\def\e{\mathsf e}
\def\bfa{\mathbf{a}}
\def\bfr{\mathbf{r}}
\def\bfk{\mathbf{k}}
\def\ZO{\{0,1\}}

\newcommand{\unity}{\ensuremath{\mathbbm 1}}

\theoremstyle{definition}
\newtheorem{theorem}{Theorem}
\newtheorem{lemma}[theorem]{Lemma}

\begin{document}
\title{Seedless extractors for device-independent quantum cryptography}

\author{Cameron Foreman}
\email{cameron.foreman@quantinuum.com}
\affiliation{Quantinuum, Partnership House, Carlisle Place, London SW1P 1BX, United Kingdom}
\affiliation{Department of Computer Science, University College London, United Kingdom}

\author{Llu\'is Masanes}
\affiliation{Department of Computer Science, University College London, United Kingdom}
\affiliation{London Centre for Nanotechnology, University College London, United Kingdom}

\begin{abstract}
  \noindent
  Device-independent (DI) quantum cryptography aims at providing 
  secure cryptography with minimal trust in, or characterisation of, the underlying quantum devices. 
  A key step in DI protocols is randomness extraction (or privacy amplification), which typically requires a \textit{seed} of additional bits with sufficient entropy and statistical independence from any bits generated during the protocol. 
  In this work, we propose a method for extraction in DI protocols that does not require a seed and is secure against computationally unbounded quantum adversaries. The core idea is to use the Bell violation of the raw data, rather than its min-entropy, as the extractor promise. We present a complete security proof in a model where the experiment uses memoryless measurement devices acting on an arbitrary joint (across all rounds) state. Our results mark a first step in this alternative, seedless, approach to extraction in DI protocols.
\end{abstract}
\maketitle
\section{Introduction}

\noindent
Device-independent (DI) quantum cryptography has emerged as the gold standard in terms of security for cryptographic applications \cite{acin2007device, pironio2009device}. 
This paradigm provides security with minimal trust in, or characterisation of, the underlying quantum devices.
This is achieved by exploiting quantum non-locality, that is, correlations which violate Bell inequalities \cite{bell1964einstein}.
Some applications of DI quantum cryptography include key distribution \cite{mayers1998quantum, renner2008security, vazirani2019fully, nadlinger2022experimental, zapatero2023advances}, randomness expansion \cite{colbeck2011private, acin2016certified}, randomness amplification \cite{colbeck2012free}, and many more \cite{gheorghiu2015robustness, kundu2022device, broadbent2023device}. 

A crucial step in numerous DI protocols is that of \textit{randomness extraction} (similarly \textit{privacy amplification}),
which involves generating a near-perfect random output (a \textit{final output}) by classically processing some imperfect, somewhat random input (a \textit{extractor input}) derived from measurement outcomes.
To date, randomness extraction in DI tasks has {required} the consumption of a \textit{seed} of bits that must, at a minimum, be sufficiently statistically independent of the quantum hardware \cite{arnon2015quantum, ball2022randomness} and sufficiently random from the adversary's perspective \cite{li2023two}. {In practice, this requirement is often more stringent, such as {requiring} a perfectly uniform and independent seed \cite{ma2013postprocessing, cryptomite}.}

In this work, we demonstrate that randomness extraction in DI quantum cryptography can be achieved using a deterministic algorithm (i.e., without a seed), while preserving security against computationally unbounded quantum adversaries -- and, by extension, against computationally unbounded classical adversaries.
The intuition behind our results is that the violation of Bell inequalities not only guarantees a lower bound on the min-entropy of the outcomes, but certain statistical independence between the outcomes of different rounds of the experiment. 
This concept can be understood from results in \textit{self-testing}, such as those {showing} that maximal violation of the CHSH inequality \cite{clauser1969proposed} implies the measured bipartite state is essentially pure \cite{wang2016all}, indicating {that there is} no correlation between rounds.

Prior to this work, only the min-entropy promise has been exploited in DI protocols, which necessitates the use of randomised (seeded or multi-source) extractors.
{It is shown in \cite{santha1986generating} that it is impossible to deterministically extract even a single bit from an input characterised solely by min-entropy.
However, our approach involves designing extractors that take advantage of the promise of Bell violation instead.}
This stronger promise allows us to {eliminate} the need for a seed. 
{In our proofs, we focus on the CHSH inequality due to its simple experimental setup (a quantum device with two isolated parts) and widespread use in DI protocols, although our proof techniques can be readily applied to other Bell inequalities.}

{The results of this paper are proven for the case where the experimental state can be arbitrary (e.g., with correlations across rounds), while the two measurement devices are memoryless (or equivalently, where each round's measurement is performed using a separate, non-communicating device).}
Although not fully general, they mark an important step in initiating a new \textit{seedless} approach to randomness extraction in DI quantum cryptography protocols, with numerous problems to {explore} (see `Conclusion and discussion' in \Cref{sec:conclusions}).
From a fundamental perspective, by exploiting the full power of Bell inequality violation, we identify a new class of distributions that can be both deterministically extracted from and generated by a realisable experimental process.
This contributes to a long line of research in computer science exploring deterministic randomness extraction, e.g. \cite{von1963various, gabizon2008deterministic, gabizon2006deterministic, kamp2006deterministic, li2016improved, chattopadhyay2016extractors, beigi2018optimal} (see \cite{chattopadhyay2022recent} for an overview).  
Moreover, we observe that arbitrary violation of the CHSH Bell inequality is sufficient for seedless extraction -- which possibly speaks to the power of Bell non-locality as a resource \cite{buhrman2010nonlocality, masanes2009universally, gallego2013full, wooltorton2023device, farkas2023unbounded}.

{It is important to note that DI protocols include a step to test the degree of Bell violation, which requires random numbers to choose the measurement settings in each round. Additionally, several other subroutines rely on randomness; for example, key distribution protocols need preshared randomness for public channel authentication.}
Therefore, the seedless extraction we present in this work does not {completely} eliminate the need for initial randomness.
However, we expect that the initial randomness required for a Bell test must satisfy weaker statistical conditions than that for both a Bell test and seeded extraction.
Thus, we {hope} that future contributions using the techniques of this work will improve the capabilities of DI protocols, particularly randomness amplification \cite{gallego2013full, ramanathan2016randomness, ramanathan2018practical, kessler2020device, ramanathan2021no, foreman2023practical}\@.

In \Cref{sec:results} we review the relationship between Bell-inequality violation and randomness, we introduce the notation and main definitions of the paper, and prove our two theorems on seedless extraction.
In \Cref{sec:estimation} we apply these results to a spot-checking based DI protocol, in which we show that our seedless extractors allow for extraction in the case of arbitrary low CHSH violation,
and have unity rate in the limit of maximal CHSH violation.
In \Cref{sec:conclusions} we conclude and list the most important open problems on DI seedless extractors.
The Appendix contains the proof of two lemmas and a theorem.

\section{Device-independent seedless extractors} 
\label{sec:results}

\subsection{Randomness and violation of Bell inequalities}

\noindent
Suppose Alice has a quantum system with Hilbert space $\A$, which can be measured with two observables {labelled} $x\in\ZO$ with outcomes $a\in\ZO$ represented by the positive operator-valued measure (POVM) elements $A(a|x)$. 
Analogously, Bob has a system $\B$ and two observables $y\in\ZO$ with outcomes $b\in\ZO$ and POVMs $B(b|y)$.
The joint state of $\A\otimes \B$ is denoted by $\rho_{\A\B}$.
With this notation, we can write the CHSH inequality \cite{clauser1969proposed} as
\begin{align}\label{Bell operator}
  \mathrm{CHSH}
  = \!\sum_{x,y,a,b=0}^1 \!\!
  \tr\! \left[\rho_{\A\B} A(a|x) B(b|y) \right] 
  (-1)^{a+b+xy} \leq 2\ .
\end{align}
If this inequality is violated then no locally causal model can explain the observed correlations \cite{clauser1969proposed}.
{Note that, in our notation, the} operator $A(a|x)$ is meant to act trivially on $\B$, so we can write $A(a|x) B(b|y)$ instead of $A(a|x)\otimes B(b|y)$.

The predictability of outcome $a$ when measuring $x$ can be quantified by the bias of the probability distribution of $a$ in the following sense
\begin{align}
  \big|\P(a=0|x)-\P(a=1|x)\big|
  =
  \big|\tr\! \left(\rho_{\A\B} \left[A(0|x)-A(1|x)\right] \right)\big|\ .
\end{align}
The following theorem proven in \cite{masanes2011secure} tells us that the stronger the CHSH violation, the less predictable the outcome $a$.

\begin{theorem} \label{thm:operator-inequality}
For any pair of Hilbert spaces $\A, \B$ and measurements $\{A(0|x), A(1|x)\}$ on $\A$ and $\{B(0|y), B(1|y)\}$ on $\B$,
we define the shifted CHSH operator as
\begin{align}\label{def:S}
  S= \mu_s \unity - \nu_s \!\!\!
  \sum_{x,y,a,b=0}^1 \!\! A(a|x) B(b|y)(-1)^{a+b+xy} \ ,
\end{align}
with coefficients
\begin{align}
  \mu_s = 2\left(2 -\frac{s^2} 4\right)^{\!\!-1/2},
  \qquad
  \nu_s = \frac s 4  \left(2 -\frac{s^2} 4 \right)^{\!\!-1/2}\ .
\end{align}
The following two semi-definite inequalities
\begin{align} \label{eq:G} 
  \pm \left[A(0|0)-A(1|0)\right]
  \leq  S\ , 
\end{align} 
hold for all $s \in (2, 2\sqrt{2})$.
\end{theorem} 

\noindent {The shifted CHSH operator~\eqref{def:S} includes an identity term with a positive coefficient and the CHSH operator from~\eqref{Bell operator} with a negative coefficient. This specific parametrisation, defined using $s$, is known to yield a tight family of inequalities when $s$ corresponds to the CHSH violation. 
Notably, inequality~\eqref{eq:G} indicates that greater CHSH violation corresponds to reduced predictability, since
\begin{align}\label{predict-bound}
  \left|\tr\! \left(\rho_{\A\B} \left[A(0|0)-A(1|0)\right] \right)\right|
  \leq
  \tr\! \left(\rho_{\A\B} S \right)\ .
\end{align}
This fact, along with its generalisation to other Bell inequalities, is central to DI quantum cryptography and crucial for the results in this section. Our proofs rely on having a family of semi-definite inequalities which relate the predictability of generated outcomes to Bell inequality violation. Since \Cref{predict-bound} is known to be tight and \Cref{def:S} has a simple form, we focus on the CHSH inequality when presenting our results. However, our proof can be easily adapted to other Bell inequalities.}

\subsection{Security in the presence of a quantum adversary} 

\noindent
The process for generating the {extractor input} $\a = (a_1,\ldots, a_{n_\r})$ consists of $n_\r$ rounds labelled by $i\in \{1,\ldots, n_\r \}$. In round $i$ Alice performs the measurement $\{A_i(0|0), A_i(1|0)\}$ on the system with Hilbert space $\A_i$, and obtains the outcome $a_i$. {Specifically, this means that when Alice is generating the {extractor input}, she uses the input $x_i = 0$ in every round.
Similarly, in round $i$ Bob has a system with Hilbert space $\B_i$, but our protocol does not require Bob to make measurements for {extractor input} generation.}
The adversary (Eve) holds an arbitrary quantum system with Hilbert space $\E$.
We use the notation $\A_\r = \bigotimes_{i=1}^{n_\r} \A_i$ and $\B_\r = \bigotimes_{i=1}^{n_\r} \B_i$ and understand that the action of $A_i(a|x)$ on $\A\otimes \B$ is trivial on all factors but $\A_i$. 
The factorisation of the total Hilbert space $\A_\r \otimes \B_\r \otimes \E$ enforces the assumption of no-signalling between Alice, Bob and Eve.
Additionally, the fact that we model every round $i$ with a different Hilbert space $\A_i \otimes \B_i$ enforces the assumption that devices have no memory. 

The global state shared among Alice, Bob and Eve is $\rho_{\A_\r \B_\r \E}$, and the reduced state of Alice and Eve $\rho_{\A_\r \E}$. 
The {final output} $\k= (k_1, \dots, k_m) \in \ZO^m$ is produced by applying the (deterministic) function $g:\{0,1\}^{n_\r} \to \{0,1\}^m$ to the {extractor input} $\a \to \k = g(\a)$. In the following sections we characterise the functions $g$. 
Although the {final output} $\k$ is a classical system, it is convenient to associate to it a Hilbert space $\K = \mathbb C^{2^m}$ and to represent its values by an orthonormal basis $\ket {\bf k}\in \K$\@.
After Alice measures all her systems $\A_i$ {for $i \in \{1, \ldots, n_{\r}\}$} and generates the {final output on system} $\K$, the joint state of systems $\K \otimes \E$ is
\begin{align}\label{def:rho KE}
  \rho_{\K\E} = 
  \sum_{\k\in \{0,1\}^m} 
  \sum_{\a\in \{0,1\}^{n_{\r}}} \!\!
  \ket{\k}\!\!\bra\k _\K
  \delta_{g(\a)}^\k 
  \tr_{\A_\r}\!\left[\rho_{\A_\r \E}\prod_{i=1}^{n_{\r}} A_i(a_i|0)\right]\ .
\end{align}  
{We include the labels $\K,\E,\A_\r$ to the operators to specify on which Hilbert space they act, hence we don't need to use the symbol $\otimes$.} The goal of Alice and Bob is to produce a state $\rho_{\K\E}$ that is indistinguishable from an ideal {final output} $u_\K\rho_{\E}$, where we define the uniform state (sometimes called maximally mixed) as
\begin{align}
  u_\K = 2^{-m} 
  \!\!\!\!\sum_{\k\in \{0,1\}^m}\!\! 
  \ket{\k}\!\! \bra{\k}_\K\ .
\end{align}
This indistinguishability can be formalised as a bound on the trace norm 
\begin{align}\label{extractor error}
  \left\| \rho_{\K\E} - u_\K \rho_{\E} \right\|_1 
  \leq \epsilon\ ,  
\end{align}
which is defined as $\|M\|_1 = \tr\sqrt{M^\dagger M}$ for any operator $M$.
The bound in \Cref{extractor error} implies that any cryptographic task which requires an ideal final output $u_\K \rho_{\E}$ as a resource, is also secure if performed using the real {final output} $\rho_{\K\E}$, up to an error of probability $\epsilon$.
Hence, we say that our seedless extraction protocol can be composed with any other cryptographic protocol: it is universally composable \cite{canetti2001universally}.

\subsection{XOR is a seedless extractor}

\noindent
The XOR function allows to extract a single bit $k\in \{0,1\}$ with an error that can be made exponentially small in $n_\r$.
The computational cost of XOR is $O(n_{\r})$, showing that seedless extractors {do not need to} be {computationally} hard to implement.

\begin{theorem}[XOR]\label{thm:xor}
  After measuring the $n_\r$-round state $\rho_{\A_\r \B_\r \E}$ with the observables $\{A_i(a_i|0):a_i=0,1\}$ for all rounds $i=1, \ldots, n_\r$ and applying the XOR function 
  \begin{align} \label{fun:xor}
    g({\bf a}) = \sum_{i=1}^{n_{\r}} a_i \bmod 2
    \in \ZO 
  \end{align}
  to the outcomes $k= g(a_1, \ldots, a_{n_\r})$, the resulting state $\rho_{\K\E}$ written in \Cref{def:rho KE} satisfies
  \begin{align}\label{Th 2}
    \left\| \rho_{\K \E} - u_{\K}\rho_{\E} \right\|_1
    \leq  
    \tr\!\left[\rho_{\A_\r \B_\r} \prod_{i=1}^{n_\r} S_i\right]\ ,
  \end{align}
  for all $s\in [2,2\sqrt 2]$.
\end{theorem}  

\noindent
The above result shows that the larger the violation of CHSH {(i.e. the smaller the expectation of $\prod_i S_i$)}, the smaller the distance between the real and the ideal {final output}s.

\begin{proof}\label{proof-xor}
  We start by defining the operator
  \begin{align}
    C_i = A_i(0|0)-A_i(1|0)\ ,
  \end{align}
  and noting that
  \begin{align}\label{def:A(a|x)}
    A_i(a_i|0) = \frac 1 2 \left(\unity + (-1)^{a_i} C_i \right)\ .
  \end{align} 
  As proven in \cite{gelfand1943imbedding}, there is no loss of generality in assuming that the operators $A_i(a_i |0)$ are projectors, which implies that $C_i$ is full-rank.

  Next, we substitute \Cref{def:A(a|x)} into the joint state after Alice generates the {final output}~\eqref{def:rho KE} and expand the product $\prod_i \left(\unity +(-1)^{a_i} C_i\right)$ into $2^{n_\r}$ terms labelled by the vectors ${\bf r} \in \{0,1\}^{n_\r}$,
  \begin{align}
    \nonumber
    \rho_{\K \E}
    &= \sum_{k,\mathbf{a}} \ket k \!\! \bra k_\K
    \delta_{g(\mathbf{a})}^{k} \tr_{\A_\r}\!\left[ \rho_{\A_\r \E} \prod_i \frac 1 2 \left(\unity +(-1)^{a_i} C_i\right) \right] 
    \\ \label{eq:60}
    &= \sum_{k,\mathbf{a}} \ket k \!\! \bra k_\K \delta_{g(\mathbf{a})}^{k} \tr_{\A_\r}\!\left[ \rho_{\A_\r \E}\, 2^{-n_\r} \sum_{\bf r} \prod_i (-1)^{a_i r_i} C_i^{r_i} \right]\ ,
  \end{align} 
  where we have used the power identities $C_i^0 = \unity$ and $C_i^1 = C_i$ for full-rank operators.
  Next, we write the XOR function as a scalar product $g({\bf a}) = {\bf a \cdot 1} \bmod 2$ with the vector ${\bf 1} = (1,\ldots, 1)$. This allows us to write the Kronecker delta as
  \begin{align}
    \delta_{g(\mathbf{a})}^{k} =
    \frac 1 2 \left[1+(-1)^{{\bf a\cdot 1}+k} \right]\ ,
  \end{align}
  and perform the summation
  \begin{align}
    \nonumber
    2^{-n_\r} \sum_{\bf a} \delta_{g(\mathbf{a})}^{k} (-1)^{\bf a\cdot r}
    &=
    2^{-n_\r -1} \sum_{\bf a} \left[ (-1)^{\bf a\cdot r} + (-1)^{{\bf a\cdot (r-1)}+k} \right]
    \\ \label{eq:16} &=
    2^{-1} \left[ \delta_{\bf r}^{\bf 0} 
    +(-1)^{k} \delta_{\bf r}^{\bf 1}\right]\ ,
  \end{align}
  where ${\bf 0} = (0,\ldots, 0)$.
  Substituting \Cref{eq:16} into the state~\eqref{eq:60} gives 
    \begin{align}
    \rho_{\K \E}
    &= \sum_{k} \frac 1 2 \ket k \!\! \bra k_\K
    \left(\rho_\E + (-1)^k\tr_{\A_\r}\!\left[ \rho_{\A_\r \E}\, \mbox{$\prod_i$} C_i \right]\right),
  \end{align} 
  which then can be used in the trace norm yielding 
  \begin{align}  
    \nonumber
    \left\| \rho_{\K \E} - u_{\K} \rho_{\E} \right\|_1
    &= \sum_{k} \frac 1 2 \left\| (-1)^k \tr_{\A_\r} \!\left( \rho_{\A_\r \E} \mbox{$\prod_i$} C_i \right) \right\|_1 
    \\ \label{eq:26}
    &= \left\| \tr_{\A_\r }\!\left(\rho_{\A_\r \E}\,\mbox{$\prod_i$} C_i\right)\right\|_1 
  \end{align}
  For any Hermitian operator $X$, its trace norm satisfies $\|X\|_1 = \max_H \tr(HX)$, where $H$ is constrained to be Hermitian and have eigenvalues $\pm1$.
  Therefore, there is an Hermitian operator $H$ acting on $\E$, with spectrum $\pm1$, which satisfies
  \begin{align}
     \left\|\tr_{\A_\r}\!\left( \rho_{\A_\r \E}
     \mbox{$\prod_i$} C_i\right) \right\|_1 
     =
     \tr\!\left( \rho_{\A_\r \E}
     \left[\left( \mbox{$\prod_i$} C_i \right) H\right] \right)\ .
  \end{align}
  We can write the spectral decomposition $H = H_+ -H_-$ with some projectors $H_\pm$ on $\E$ and obtain
  \begin{align}\label{eq separat}
     \left\|\tr\!\left( \rho_{\A_\r \E}\,
     \mbox{$\prod_i$} C_i\right) \right\|_1 
     =
     \tr\!\left( \rho_{\A_\r \E}
     \left[ \mbox{$\prod_i$} C_i \right] H_+ \right)
     -     
     \tr\!\left( \rho_{\A_\r \E}
     \left[ \mbox{$\prod_i$} C_i \right] H_- \right)\ .
  \end{align}
  At this point we recall \Cref{thm:operator-inequality}, which can be written as $\pm C_i \leq S_i$.
  Note that here, operator $C_i$ acts trivially on $\B_i$ while $S_i$ does not.
  Using \Cref{lem:product SC} we can generalise \Cref{thm:operator-inequality} to $\pm \mbox{$\prod_i$} C_i \leq \mbox{$\prod_i$} S_i$, which implies $\pm [\mbox{$\prod_i$} C_i ] H_{\pm} \leq [\mbox{$\prod_i$} S_i] H_{\pm}$ and 
  \begin{align}
     \pm \tr\!\left( \rho_{\A_\r \E}
     \left[ \mbox{$\prod_i$} C_i \right] H_\pm \right)
     \leq
     \tr\!\left( \rho_{\A_\r \B_\r \E}
     \left[ \mbox{$\prod_i$} S_i\right] H_\pm \right).
  \end{align}
  Finally, substituting this in \Cref{eq separat} and using the fact that $H_+ +H_- = \unity$, we obtain
  \begin{align}
     \left\|\tr_{\A_\r }\!\left( \rho_{\A_\r \E}\,
     \mbox{$\prod_i$} C_i\right) \right\|_1 
     \leq
     \tr\!\left( \rho_{\A_\r \B_\r}
     \mbox{$\prod_i$} S_{i} \right)\ ,
  \end{align}
  which concludes the proof.
\end{proof} 

\subsection{Seedless extractors with arbitrary output length}

\noindent
In this section, we analyse seedless extractors {that} produce {an extractor output} $\k$ of arbitrary length $m$. 
The proof relies on randomised methods{, meaning} we do not obtain explicit constructions and the resulting extractors are likely computationally hard to implement. {In future work, we will show that certain linear functions are seedless extractors and can be implemented in computation time $O(n_\r \log n_\r)$ \cite{foreman-in-prep}.}
The following lemma is proven in \Cref{proof-existence}.

\begin{lemma}\label{lem:balanced functions}
    {There exist functions $g: \{0,1\}^{n_{\r}} \rightarrow \{0,1\}^m$, for $n_{\r} > 5$ and $n_{\r} - m > 0$, satisfying
    \begin{align}
        \label{sum-eval}
        \left| 
        \sum_{\bfa} \left( \delta_{g(\mathbf{a})}^{\bf k} - 2^{-m} \right) (-1)^{\bf a\cdot r} \right|
        \leq  n_{\r}^2\sqrt{2}^{n_{\r} - m}\ ,
    \end{align} 
    for all $\mathbf{k} \in \{0,1\}^{m}$ and all $\mathbf{r} \in \{0,1\}^{n_{\r}}$.} We call such functions \textit{$m$-bit extractor functions}.
\end{lemma}

\noindent {This property is crucial for proving the following theorem, as it enables us to bound each coefficient after expanding the trace distance between the real and ideal output using the triangle inequality. Specifically: (1) For $\mathbf{r} = \mathbf{0}$, this property ensures that the sizes of the sets $\{\mathbf{a} \mid g(\mathbf{a}) = \mathbf{k}\}$ are similar for all $\mathbf{k}$. (2) For $\mathbf{r} \neq \mathbf{0}$, it ensures that the sizes of the sets $\{\mathbf{a} \mid g(\mathbf{a}) = \mathbf{k} \text{ and } \mathbf{a} \cdot \mathbf{r} = 0\}$ and $\{\mathbf{a} \mid g(\mathbf{a}) = \mathbf{k} \text{ and } \mathbf{a} \cdot \mathbf{r} = 1\}$ are similar for all $\mathbf{k}$.}

\begin{theorem}\label{thm-existence}
  Let $g: \{0,1\}^{n_{\r}} \rightarrow \{0,1\}^m$ be a {function satisfying the condition~\eqref{sum-eval}.}
  After measuring the $n_{\r}$-round state $\rho_{\A_\r \B_\r \E}$ with the observables $\{A_i(a_i|0):a_i=0,1\}$ for all $i=1, \ldots, n_{\r}$ and applying the function $\mathbf{k} = g(\mathbf a)$ to the outcomes $\mathbf a = (a_1, \ldots, a_{n_{\r}})$, the resulting state $\rho_{\K\E}$ written in \Cref{def:rho KE} satisfies
  \begin{align}\label{Lem-rand}
    {\| \rho_{\K \E} - u_{\K} \rho_{\E} \|_1
    \leq 
    n_{\r}^2 \sqrt{2}^{m-n_{\r}} \tr\!\left[ \rho_{\A_\r \B_\r} \prod_{i=1}^{n_{\r}} 
    (\unity +S_{i})
    \right]\ ,}
  \end{align}
  for all $s\in [2,2\sqrt 2]$.
\end{theorem} 

\noindent
{The above result shows that, the smaller the expectation of $\prod_i (\unity +S_i)$ (i.e. the larger the violation of CHSH), the smaller the distance between the real and the ideal {final output}s.}
If we fix this distance to a specific value $\| \rho_{\K \E} - u_{\K} \rho_{\E} \|_1  =\epsilon$, then the larger the violation of CHSH, the larger the length $m$ of the {extractor output}.

\begin{proof}
  We start by substituting \Cref{def:A(a|x)} in the joint state after Alice generates the {final output}~\eqref{def:rho KE} and expanding the product $\prod_i \left(\unity +(-1)^{a_i} C_i\right)$ into $2^{n_{\r}}$ terms labelled by the vectors ${\bf r} \in \{0,1\}^{n_{\r}}$,
  \begin{align}
    \nonumber
    \rho_{\K \E}
    &= \sum_{\mathbf k,\mathbf{a}} 
    \ket{\mathbf k} \!\! \bra{\mathbf k}_\K
    \delta_{g(\mathbf{a})}^{\mathbf{k}} \tr_{\A_\r}\!\!\left[ \rho_{\A_\r \E} \prod_i \frac 1 2 \left(\unity +(-1)^{a_i} C_i\right) \right] 
    \\ \label{eq:60c}
    &= \sum_{\mathbf k,\mathbf{a}} 
    \ket{\mathbf k} \!\! \bra{\mathbf k}_\K
    \delta_{g(\mathbf{a})}^{\mathbf{k}} \tr_{\A_\r}\!\!\left[ \rho_{\A_\r \E}\, 2^{-n_{\r}} \sum_{\bf r} \prod_i (-1)^{a_i r_i} C_i^{r_i} \right]\ ,
  \end{align} 
  where we have used the power identities $C_i^0 = \unity$ and $C_i^1 = C_i$ for full-rank operators. 
  {After substituting this in the left-hand side of \Cref{Lem-rand}, applying the triangular inequality and using promise~\eqref{sum-eval}, we obtain
  \begin{align} 
    \nonumber
    \left\| \rho_{\K \E} - u_{\K} \rho_{\E} \right\|_1 
    &= 
    2^{-n_{\r}} \sum_{\mathbf k}  
    \left\| \sum_{\mathbf{r}} \left( 
    \sum_{\bfa} \left( \delta_{g(\mathbf{a})}^{\bf k} - 2^{-m} \right) (-1)^{\bf a\cdot r} \right)
    \tr_{\A_\r}\!\!\left[ \rho_{\A_\r \E}
    \prod_i C_i^{r_i}\right] \right\|_1 \\
    \nonumber &\leq 
     2^{-n_{\r}} \sum_{\mathbf k}
     \sum_{\mathbf{r}} 
     \left| \sum_{\mathbf{a}} \left(\delta_{g(\mathbf{a})}^{\bf k} - 2^{-m} \right) (-1)^{\bf a\cdot r}
     \right| 
     \left\| 
     \tr_{\A_\r}\!\!\left[ \rho_{\A_\r \E}
     \prod_i C_i^{r_i}\right] \right\|_1 
     \\ \nonumber &\leq 
     2^{-n_{\r}} \sum_{\mathbf{k}}
     \sum_{\mathbf{r}} 
     n_{\r}^2 \sqrt{2}^{n_{\r}-m} 
     \left\| 
     \tr_{\A_\r}\!\! \left[ \rho_{\A_\r \E}
     \prod_i C_i^{r_i}\right] \right\|_1 
     \\ \label{eq A9} &=        
     n_{\r}^2\sqrt{2}^{m-n_{\r}}
     \sum_{\mathbf{r}} 
     \left\| 
     \tr_{\A_\r}\!\! \left[ \rho_{\A_\r \E}
     \prod_i C_i^{r_i}\right] \right\|_1\ .
  \end{align}}
  Following the same steps as in the proof of \Cref{thm:xor}: there are two complementary projectors $H_\pm$ acting on $\cal E$ such that 
  \begin{align}\label{eq A11}
     \left\|\tr\!\left( \rho_{\A_\r \E}\,
     \mbox{$\prod_i$} C_i^{r_i}\right) \right\|_1 
     =
     \tr\!\left( \rho_{\A_\r \E}
     \left[ \mbox{$\prod_i$} C_i^{r_i} \right] H_+ \right)
     -     
     \tr\!\left( \rho_{\A_\r \E}
     \left[ \mbox{$\prod_i$} C_i^{r_i} \right] H_- \right)\ .
  \end{align}
  \Cref{lem:product SC} allows us to generalise \Cref{thm:operator-inequality} to
  \begin{align}
    \pm \mbox{$\prod_i$} C_i^{r_i} 
    \leq 
    \mbox{$\prod_i$} S_i^{r_i} \ ,
  \end{align}
  for any vector $\bf r$, which in turn implies 
  \begin{align}
     \pm \tr\!\left( \rho_{\A_\r \E}
     \left[ \mbox{$\prod_i$} C_i^{r_i} \right] H_\pm \right)
     \leq
     \tr\!\left( \rho_{\A_\r \B_\r \E}
     \left[ \mbox{$\prod_i$} S_i^{r_i}\right] H_\pm \right)\ .
  \end{align}
  Substituting this in \Cref{eq A11} and using the fact that $H_+ +H_- = \unity$, we obtain
  \begin{align}
     \left\|\tr_{\A_\r}\!\left( \rho_{\A_\r \E}\,
     \mbox{$\prod_i$} C_i^{r_i}\right) \right\|_1 
     \leq
     \tr\!\left( \rho_{\A_\r \B_\r}
     \mbox{$\prod_i$} S_{i}^{r_i} \right)\ .
  \end{align}
  Finally, we substitute this in \Cref{eq A9} and conclude the proof
  {\begin{align}
    \nonumber
    \| \rho_{\K \E} - u_{\K} \rho_{\E} \|_1
    &\leq        
    n_{\r}^2 \sqrt{2}^{m-n_{\r}}
    \mbox{$\sum_{\mathbf{r}}$} 
    \tr\!\left( \rho_{\A_\r \B_\r}
    \mbox{$\prod_i$} {S}_{i}^{r_i} \right)
    \\ & =
    n_{\r}^2 \sqrt{2}^{m-n_{\r}}
    \tr\!\left( \rho_{\A_\r \B_\r}
    \mbox{$\prod_i$} \left[\unity +{S}_{i}\right] \right)\ .
  \end{align}}
\end{proof}

\section{Estimation} \label{sec:estimation}
\noindent
\Cref{thm:xor} and \Cref{thm-existence} provide a relationship between the error $\epsilon$, the length of the {final output} $m$, and the Bell-inequality violation quantified by $\langle\prod_i S_i\rangle$ or $\langle\prod_i (\unity +S_i)\rangle$.
This Bell-violation quantifier is different than the one used with standard seeded extractors, 
that is $\langle\sum_i S_i \rangle$.
For large $n$, the statistical fluctuations of $\sum_i S_i$ are small, which allows us to relate the average $\langle\sum_i S_i \rangle$ to the particular value of $\sum_i S_i$ corresponding to the estimation data $(a_j, b_j, x_j, y_j)$.
Unfortunately, the quantities $\prod_i S_i$ and $\prod_i (\unity +S_i)$ appearing in our bounds have strong fluctuations and cannot be bounded with the usual techniques.\footnote{{We note that if the experimental state is product across the rounds, the estimator simplifies significantly and can be computed using standard techniques.}} 

In this section, we present a proof technique for bounding $\langle\prod_i (\unity +S_i)\rangle$ and $\langle\prod_i S_i\rangle$ with the estimation data. For this, we use the \textit{spot-checking} procedure frequently used in DI protocols and, therefore, widely applicable. In this procedure, we randomly select a subset of rounds used for the estimation of the Bell violation, the rest of {the} rounds are used to generate the {extractor input}. This random selection limits the malicious behaviour of the devices.  
{In order to implement this procedure and prove security, the required assumptions are: 
\begin{itemize}
    \item The devices and adversary operate according to quantum theory.
    \item The classical computer used for processing and statistical analysis is trusted and functions correctly.
    \item The quantum device comprises two isolated parts that do not exchange information during each round of the experiment (i.e., they are no-signaling).
    \item The quantum device does not signal to the adversary. 
    \item The measurement devices are memoryless, meaning that the measurements in each round act on separate Hilbert spaces.
\end{itemize}
Importantly, (1) we make no assumptions about the state, which can be arbitrary and may exhibit correlations between rounds, and (2) although the measurements are modelled to act on a separate Hilbert space in each round, they do not need to be identical.}

\subsection{SPOT-CHECKING PROTOCOL FOR XOR EXTRACTION}
\label{subsec:xor_prot}
\noindent
In what follows, we provide a complete description of this estimation protocol and show that our XOR seedless extractor is able to extract a bit with arbitrary small error. 

{ \sf 
\begin{description}
\item[Set parameters:] $n$, the total number of rounds, $p_\e \in (0, 1)$, the probability for a round to be used for estimation, and $\epsilon>0$, the tolerable error.

\item[Data generation.] For each round $l\in \{1,\dots, n\}$ repeat the following steps:
  \begin{enumerate}
    \item Generate the random variable $t_l\in \{\mathtt{estimation}, \mathtt{rawbit}\}$ with probabilities $p_\e$ and $p_\r= 1-p_\e$ respectively.
    \item If $t_l = \mathtt{estimation}$ then: 
    \begin{enumerate}
      \item Generate the random variables $x_l, y_l\in \ZO$ with uniform distribution $\P(x_l,y_l) = 1/4$,
      \item Perform the bi-local measurement $A_l (a_l|x_l) B_l (b_l|y_l)$ with outcomes $a_l, b_l\in \ZO$,
      \item Record the variable $z_l = a_l+b_l+x_l y_l \bmod 2$, which will be used to evaluate the CHSH inequality.
    \end{enumerate}
    \item If $t_l = \mathtt{rawbit}$ then perform the local measurement $A_l (a_l|0)$ and keep the outcome $a_l$ as part of the {extractor input}.
  \end{enumerate}
  
\item[Data processing.] Denote by $n_\e \in \{0, n\}$ the number of rounds $l$ with $t_l = \texttt{estimation}$, assign an index $j\in \{1,\ldots, n_\e\}$ to each of them, and compile the estimation data $\z = (z_1\ldots, z_{n_\e})$.
  Denote by $n_\r \in \{0, n\}$ the number of rounds $l$ with $t_l = \mathtt{rawbit}$, assign an index $i\in \{1,\ldots, n_\r\}$ to each of them, and compile the {extractor input} $\a =(a_1,\ldots, a_{n_\r})$.
  The numbers $n_\e, n_\r$ are a function of $\t =(t_1, \ldots, t_n)$. 
\begin{enumerate}
  \item Calculate the length of the {final output} as a function of $\t, \z$ with the formula;
      \begin{align} \label{m(t,z)-xor}
        m(\t, \z) &= 
        \begin{cases}
            1 \quad \mathrm{if} \quad  
        \max\limits_{s,\alpha_0, \alpha_1, \beta}
        \sum_{j=1}^{n_\e}\alpha_{z_j} + (\beta - 1) n_\r - 2 \log_2(1/\epsilon) \geq 0 \\
            0 \quad \mathrm{otherwise}
        \end{cases},
      \end{align} where the maximisation over the parameters $s\in[2,2\sqrt 2]$ and $\alpha_0, \alpha_1, \beta \in \mathbb R$ is constrained by
  \begin{align} \label{main conditon 1 xor}
    p_{\r} \sqrt 2^{\beta-1} \left(\mu_s-4 \nu_s \right) 
    +p_{\e} \sqrt 2^{\alpha_0}
    = 1\ ,
    \\ \label{main conditon 2 xor}
    p_{\r} \sqrt 2^{\beta-1} \left(\mu_s +4 \nu_s \right) 
    +p_{\e} \sqrt 2^{\alpha_1}
    = 1\ .
  \end{align}
    \item Generate the {final output} $\k=g(\a) \in \ZO^{m(\t, \z)}$ by applying to the {extractor input} the XOR function $g:\ZO^{n_{\r}} \to \ZO^{m(\t, \z)}$, defined in \Cref{fun:xor}\@.
\end{enumerate}
\end{description}}

\noindent
{The expression for the extractor output length~\eqref{m(t,z)-xor} may seem unintuitive, but it is the most general form that allow us to prove the following theorem.}

\begin{theorem} \label{thm:est-prot}
  {For any choice of $n, p_\r, \epsilon$,} the above protocol generates a {final output} $\rho_{\K\E|\t, \mathbf z}$ satisfying the following security condition
  \begin{align}\label{th:est bound}
    \sum_{\ \t \in \{0,1\}^n}\!\!
    \sum_{\ \z \in \{0,1\}^{n_{\e}}} \!\!\! {\P}(\t, \z) 
    \left\| \rho_{\K\E|\t, \z} -
    u_{\K}  
    \rho_{\E|\t, \z} \right\|_1
    \leq  
    \epsilon\ ,
  \end{align}
  {where $\rho_{\K\E|\t,\z}$ is defined in Equation~\eqref{def:rho KE} with $m = m(\mathbf{t}, \mathbf{z})$ defined in Equation~\eqref{m(t,z)-xor} (XOR extraction protocol) and Equation~\eqref{m(t,z)} ($m$-bit extraction protocol).}
\end{theorem}

\noindent {This theorem (proven in \Cref{app:proof-thm-est}) demonstrates that the above protocol produces a secure final output. The expression for output length~\eqref{m(t,z)-xor} provides a general form with free parameters $\alpha_0$, $\alpha_1$, and $\beta$ which, importantly, allows a free parameter to be distributed across each of the factors in the error term when expanded (see Equation~\eqref{factors-xor}). Then, the maximisation constraints in~\eqref{main conditon 1 xor} and~\eqref{main conditon 2 xor} are required to ensure that specific operator identities are maintained while incorporating this additional freedom (in particular, to recover the identities in Equations~\eqref{id1} and~\eqref{id2}).}

\medskip\noindent
{Next, we} analyse the noise tolerance of the XOR extractor in the spot-checking protocol presented above. {Define} the relative frequency of the estimation outcomes as
\begin{align}
q_z = \frac {|\{z_j=z : j=1,\ldots, n_\e\}|} {n_\e} \ ,
\end{align}
for $z=0,1$. Which is related to the CHSH expression~\eqref{Bell operator} via
\begin{align}
\mathrm{CHSH} = 4(q_0-q_1) = 8q_0 - 4\ .
\end{align}
Note that, for any permutation $\sigma$ of $(z_1, z_2,\ldots, z_{n_\e})$ we have $m(\t, \sigma\z) =m(\t, \z)$, hence $m$ depends on $\z$ through $q_0$.
Also, $m$ depends on $\t$ via its relative frequency $n_\e/n$, which in the large-$n$ limit is $n_\e /n \approx p_\e$ with high probability.
Therefore, if we assume a constant error $\epsilon>0$, in the large-$n$ limit, the condition for positive yield (i.e.~$m=1$) is 
\begin{align}\label{chsh-min}
  \max\limits_{s,\alpha_0, \alpha_1, \beta}\, p_{\e}(\alpha_0 q_0+\alpha_1 q_1) + \beta p_{\r} \geq 0\ .
\end{align} 
subject to the constraints~\eqref{main conditon 1 xor} and~\eqref{main conditon 2 xor}\@. 

\begin{figure}[H]
  \centering
  \includegraphics[width=0.45\textwidth]{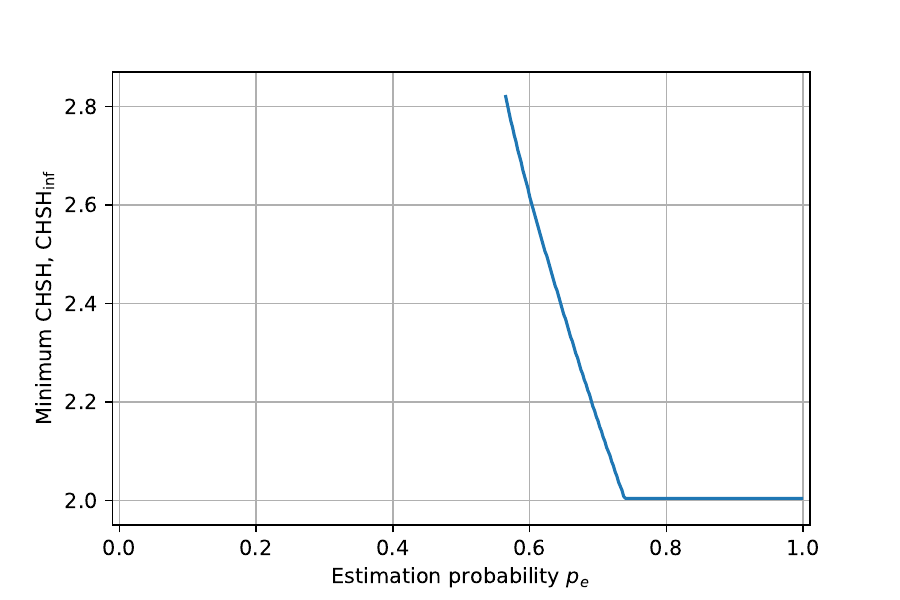}
  \caption{This plot shows the minimum CHSH value~\eqref{chsh-min} that our XOR extractor (presented in \Cref{thm:xor}) can produce a single bit with arbitrarily small error in the large-$n$ regime.
  }
  \label{fig:min-chsh}
\end{figure}

The minimum value of $\mathrm{CHSH}$ with positive yield as a function of $p_e$ is shown in \Cref{fig:min-chsh}\@. 
Note that for $p_\e \geq 0.74$ a bit can be generated with arbitrarily small Bell violation.
Interestingly, as shown by \Cref{fig:min-chsh}, a necessary requirement for the extraction of a single bit is $p_e > 0.5$. This contrasts with the fact that, in protocols with seeded extractors, only a {very small portion} is required for estimation, evidencing a limitation of our estimation method.

\subsection{SPOT-CHECKING PROTOCOL FOR $m$-BIT EXTRACTION}
\label{subsec:balanced_prot}

\noindent
In what follows, we describe the spot-checking protocol for estimating the error of seedless extraction using our $m$-bit extractor functions. 
We then calculate the maximum efficiency and extraction rates. 
The protocol is identical to that of the XOR extractor in \Cref{subsec:xor_prot} except for the {`Data processing'} step, which is replaced by the following:
\begin{description}
\item[Data processing.] Define $\epsilon, \t, n_\e, n_\r, j, i, \z, \a$ as in \Cref{subsec:xor_prot}.
\begin{enumerate}
  \item Calculate the length of the {final output} as a function of $\t, \z$ with the formula 
      {\begin{align} \label{m(t,z)}
        m(\t, \z) 
        = \left\lfloor
        \max_{s,\alpha_0, \alpha_1, \beta}
        \sum_{j=1}^{n_\e}\alpha_{z_j} + \beta n_\r -2\log_2\! \frac 1 \epsilon - 4 \log_2(n_{\r})\right\rfloor .
      \end{align}}
  where the maximisation over the parameters $s\in[2,2\sqrt 2]$ and $\alpha_0, \alpha_1, \beta \in \mathbb R$ is constrained by
  \begin{align} \label{main conditon 1}
    p_{\r} \sqrt 2^{\beta-1} \left(1 +\mu_s-4 \nu_s \right) 
    +p_{\e} \sqrt 2^{\alpha_0}
    = 1\ ,
    \\ \label{main conditon 2}
    p_{\r} \sqrt 2^{\beta-1} \left(1 +\mu_s +4 \nu_s \right) 
    +p_{\e} \sqrt 2^{\alpha_1}
    = 1\ .
  \end{align}
   
  \item Generate the {final output} $\k=g(\a) \in \ZO^{m(\t, \z)}$ by applying to the {extractor input} a function $g:\ZO^{n_{\r}} \to \ZO^{m(\t, \z)}$ {satisfying \eqref{sum-eval}}.
\end{enumerate}  
\end{description}
\noindent
{This protocol also satisfies the security condition~\eqref{th:est bound} in Theorem \ref{thm:est-prot} (proven in \Cref{app:proof-thm-est}). Similar to the XOR case, the expression for the extractor output length in~\eqref{m(t,z)} is chosen as it provides the most general form that allows us to prove security. Specifically, this form includes free parameters that can be distributed across each factor in the error term upon expansion. The maximisation constraints in~\eqref{main conditon 1} and~\eqref{main conditon 2} ensure that specific operator identities are preserved under this added flexibility. For full details, see the proof in \Cref{app:proof-thm-est}.}

\begin{figure}[H]
  \centering
  \includegraphics[width=0.44\textwidth]{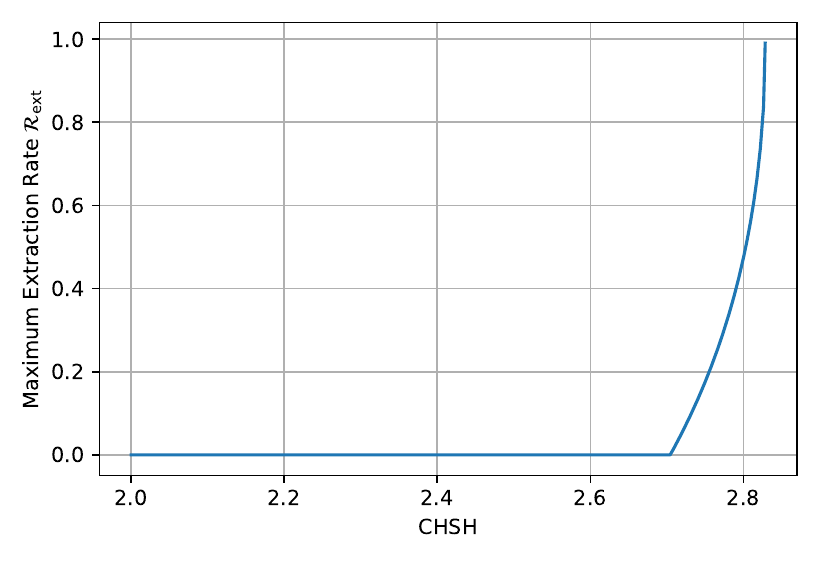}
  \includegraphics[width=0.46\textwidth]{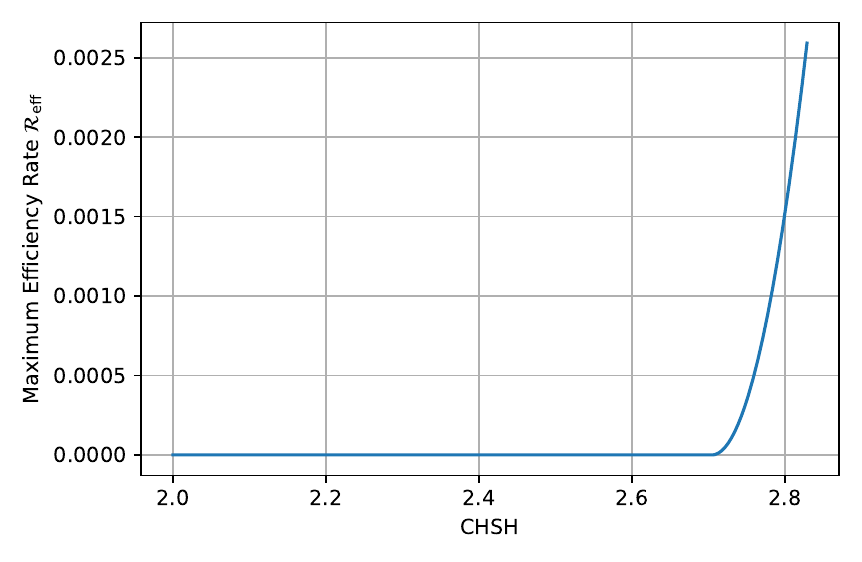}
  \caption{{The left hand side plot shows the maximum extraction rate, $ \mathcal{R}_{\mathrm{ext}}$~\eqref{ext-rate}, for different values of $\mathrm{CHSH}$. The right hand side shows the maximum efficiency rate, $\mathcal{R}_{\mathrm{eff}}$~\eqref{eff-rate}, for our $m$-bit extractor functions, for different values of $\mathrm{CHSH} = 4(q_0-q_1) \in (2, 2\sqrt{2})$. The maximisation's are performed over the variables $s \in (2, 2\sqrt{2})$, $p_{\e} \in (0,1)$ and $\alpha_0, \alpha_1, \beta \in \mathbb{R}$ such that constraints~\eqref{main conditon 1} and \eqref{main conditon 2} are satisfied.}} 
  \label{fig:rates}
\end{figure}

Using {the expression for {final output}} length~\eqref{m(t,z)} we can obtain the efficiency rate $\mathcal R _{\mathrm{eff}}$ and {the} extraction rate $\mathcal R_{\mathrm{ext}}$. 
{The extraction rate, the number of output bits per extractor input bit, is given by
\begin{align} \label{ext-rate}
  \mathcal{R}_{\mathrm{ext}} = \lim_{n\to \infty}
  \frac {m(\t,\z)} {n_\r} 
  = \frac{p_{\e}}{p_{\r}} (\alpha_0 q_0+\alpha_1 q_1) + \beta\ ,
\end{align} and the efficiency rate, the number of output bits per round, is given by
\begin{align} \label{eff-rate}
  \mathcal R_{\mathrm{eff}} = \lim_{n\to \infty}
  \frac {m(\t,\z)} n 
  = p_{\e}(\alpha_0 q_0+\alpha_1 q_1) + \beta p_{\r}\ .
\end{align} 
The maximum value of $\mathcal R_{\mathrm{ext}}$ and $\mathcal R_{\mathrm{eff}}$ as a function of CHSH is depicted in \Cref{fig:rates}. The extraction rate approaches $1$, demonstrating that our $m$-bit extractor performs optimally in the large-$n$ regime and under high CHSH violation. However, the maximum efficiency rates are very low. This is because the estimation procedure consumes a significant proportion of rounds, as with the XOR extractor.}
Therefore, independently improving techniques for estimation of our Bell value quantifier's would significantly enhance the practicality of our seedless extractors.

\section{Conclusion and discussion}
\label{sec:conclusions}
\noindent
In this work we have proven that, in the context of DI protocols, randomness extraction and privacy amplification can be done without the need of a seed.
These ideas and results constitute the first steps of a new paradigm for DI quantum cryptography, which might one day allow to minimise the resources required in various tasks.
This new paradigm contains the following open problems, which we will address in future work.
\begin{itemize}
  \item Analyse and generalise our spot-checking protocol to a DI protocol for randomness amplification. The protocol for seedless extraction presented here requires initial randomness to generate the variables $t_l$, $x_l$ and $y_l$, so that the Bell violation can be tested. However, it is not known what the minimal statistical requirements are for this initial randomness, which could be weaker than those of a Santha-Vazirani source \cite{santha1986generating} (see \cite{thinh2013bell} for necessary requirements)\@. 

  \item Characterise the seedless extractors consisting of linear functions. We have evidence that these extractors improve the efficiency rate and have a strong connection with linear error-correcting codes \cite{foreman-in-prep}\@.
  
  \item The results presented in this work are restricted to the CHSH scenario: two parties with binary inputs and outputs. However, they can be generalised to arbitrary scenarios by using the NPA hierarchy \cite{navascues2008convergent}\@. In particular, we expect that increasing the number of inputs will cause the efficiency rate to {improve} substantially, as it happens in other scenarios \cite{masanes2014full, masanes2009universally}. The efficiency rate can also be improved by using both outputs $a_l, b_l$ as the {extractor input}, or by recycling the inputs used for estimation, since these also contain randomness \cite{bhavsar2023improved}\@.
    
  \item Characterise the optimal efficiency rates of seedless extractors. There is room for improvement in several parts of our scheme, one of them being the use of the measurement outcomes of all rounds for both: estimation and {extractor input generation}. Another being the development of new proof techniques for bounding $\langle\prod_i S_i\rangle$, or related quantities.
  
  \item Our results can be implemented with only two quantum devices (one for Alice and one for Bob). However, our theorems assume that {the measurement devices} have no internal memory (as in \cite{masanes2011secure}), which is an inconvenient assumption.
  Because of the link between Bell violation and independence of the rounds discussed in the Introduction, we expect that this assumption is unnecessary. We will analyse this hypothesis in future work.
\end{itemize}

\bigskip\noindent\textbf{Acknowledgements.} 
We acknowledge useful discussions with Mafalda Almeida, Matty Hoban, Karan Khathuria, Simone Lin, Renato Renner, Ramona Wolf and Erik Woodhead. LM acknowledges financial support from the EPSRC Prosperity Partnership in Quantum Software for Simulation and Modelling
(grant EP/S005021/1).

\bibliographystyle{unsrtnat}
\bibliography{library.bib}

\appendix

\section{Proofs of Lemmas}
\label{proof-existence}

\medskip\noindent  
\textbf{\Cref{lem:balanced functions}.}
There exist functions $g: \{0,1\}^{n_{\r}} \rightarrow \{0,1\}^m$, for $n_{\r} > 5$ and $n_{\r} - m > 0$, satisfying
\begin{align}
    \label{sum-eval2}
    \left| 
    \sum_{\bfa} \left( \delta_{g(\mathbf{a})}^{\bf k} - 2^{-m} \right) (-1)^{\bf a\cdot r}  \right|
    \leq  n_{\r}^2\sqrt{2}^{n_{\r} - m}\ ,
\end{align} 
for all $\mathbf{k} \in \{0,1\}^{m}$ and all $\mathbf{r} \in \{0,1\}^{n_{\r}}$.

\begin{proof}
    {Consider a random function $G: \{0,1\}^{n_\r} \to \{0,1\}^m$, which assigns each $\bfa \in \{0,1\}^{n_\r}$ to a uniformly random and independent element $\bfk \in \{0,1\}^m$. Define $2^{n_{\r}}$ random variables indexed by $\bfa \in \{0,1\}^{n_\r}$ as
    \begin{align}
        X_{\bfa}(\bfk, \bfr) = \delta_{G(\bfa)}^{\bfk} (-1)^{\bfa \cdot \bfr}\ ,
    \end{align}
    where $\delta_{G(\bfa)}^{\bfk} = 1$ if $G(\bfa) = \bfk$ and is otherwise 0.
    For any particular value $\bfk^*$ and $\bfr^*$, these random variables are independent (since $G$ is a random function) and satisfy the probability assignment 
    \begin{align}
        &\P(X_{\bfa}(\bfk^*, \bfr^*) = 0) = (1 - 2^{-m})\ , \\
        &\P(X_{\bfa}(\bfk^*, \bfr^*) = (-1)^{\bfa \cdot \bfr^*}) = 2^{-m}\ .
    \end{align}
     Therefore, each $X_\bfa(\bfk^*, \bfr^*)$ has mean $\mathbb{E}(X_\bfa(\bfk^*, \bfr^*)) = (-1)^{\bfa \cdot \bfr^*}2^{-m}$ and second moment $\mathbb{E}(X_\bfa(\bfk^*, \bfr^*)^2) = 2^{-m}$.}
    
    {Bernstein's inequality states; for independent random variables $X_1, \ldots X_{n}$,
    \begin{align}
        \P(|\sum_{i=1}^{n} X_i - \sum_{i=1}^{n} \mathbb{E}(X_i)| \geq C) \leq 2\exp\left( \frac{-C^2}{2(\sum_{i=1}^{n} \mathbb{E}(X_i^2) + \frac{1}{3} \max\limits_i|X_i| C)}\right)\ .
    \end{align}
    Using our variables $X_\bfa(\bfk^*, \bfr^*)$, we obtain
    \begin{align}
        \nonumber \P(|\sum_{\bfa} X_\bfa(\bfk^*, \bfr^*) - \sum_{\bfa} (-1)^{\bfa \cdot \bfr^*}2^{-m} | \geq C) &\leq 2\exp\left( \frac{-C^2}{2(\sum_{\bfa}2^{-m} + \frac{1}{3}C)}\right)\\
        &= 2\exp\left( \frac{-C^2}{2(2^{n_{\r}-m} + \frac{1}{3}C)}\right)\ .
    \end{align}
    When $\bfr^* = \mathbf{0}$, $\sum_{\bfa} (-1)^{\bfa \cdot \bfr^*}2^{-m} = \sum_{\bfa} 2^{-m} = 2^{n_\r - m}$,
    and so
    \begin{align}
        \P(|\sum_{\bfa} \left(X_\bfa(\bfk^*, \bfr^* = \mathbf{0}) - 2^{- m}\right)| \geq C) &\leq 2\exp\left( \frac{-C^2}{2(2^{n_{\r}-m} + \frac{1}{3}C)}\right)\ .
    \end{align}
    When $\bfr^* \neq \mathbf{0}$, $\sum_{\bfa} (-1)^{\bfa \cdot \bfr^*}2^{-m} = 0$, 
    and so  
    \begin{align}
        \P(|\sum_{\bfa} X_\bfa(\bfk^*, \bfr^* \neq \mathbf{0})| \geq C) &\leq 2\exp\left( \frac{-C^2}{2(2^{n_{\r}-m} + \frac{1}{3}C)}\right)\ .
    \end{align}
    Together, this can be expressed for any $\bfr^* \in \{0,1\}^{n_\r}$ as
    \begin{align}
        \P(|\sum_{\bfa} \left(X_\bfa(\bfk^*, \bfr^*) - (-1)^{\bfa \cdot \bfr^*}2^{- m}\right)| \geq C) &\leq 2\exp\left( \frac{-C^2}{2(2^{n_{\r}-m} + \frac{1}{3}C)}\right)\ ,
    \end{align} 
    and, setting $C = n_{\r}^2 \sqrt{2}^{n_{\r} - m}$, leads to
    \begin{align}
        \P(|\sum_{\bfa} \left( X_\bfa(\bfk^*, \bfr^*) - (-1)^{\bfa \cdot \bfr^*}2^{- m}\right)| \geq n_{\r}^2 \sqrt{2}^{n_{\r} - m}) 
        \\
        \leq p^* :=
        2 \exp\left( \frac{-n_{\r}^4 2^{n_{\r} - m}}{2(2^{n_{\r}-m} + \frac{1}{3} n_{\r}^2 \sqrt{2}^{n_{\r} - m})}\right)\ .
    \end{align}}
    
    {Next, we apply the union bound to lower bound the probability that Equation~\eqref{sum-eval2} holds for all $\bfr$ and $\bfk$, as 
    \begin{align}
        \nonumber &\hspace{-1cm}
        \bigcap_{\bfk, \bfr} \P(|\sum_{\bfa} \left(X_\bfa(\bfk, \bfr) - (-1)^{\bfa \cdot \bfr}2^{- m}\right)| \leq C)
        \\ &= 1 - \bigcup_{\bfk, \bfr} \P(|\sum_{\bfa} \left(X_\bfa(\bfk, \bfr)- (-1)^{\bfa \cdot \bfr}2^{- m}\right)| \geq C) \\
        \nonumber &\geq 1 - \sum_{\bfk, \bfr} \P(|\sum_{\bfa} \left(X_\bfa(\bfk, \bfr) - (-1)^{\bfa \cdot \bfr}2^{- m}\right)| \geq C) \\
        \nonumber &= 1 - 2^{n_{\r}+m} \P(|\sum_{\bfa} \left(X_\bfa(\bfk^*, \bfr^*) - (-1)^{\bfa \cdot \bfr^*}2^{- m}\right) |\geq C) \\
        &\geq 1 - 2^{n_\r + m} p^*\ .
    \end{align}
    Finally, to prove the existence of functions satisfying Equation~\eqref{sum-eval2} for all $\bfk$ and $\bfr$, it suffices to show that $1 - 2^{n_{\r} + m}p^* > 0$, i.e., $2^{n_{\r} + m}p^* < 1$. Given $n_{\r} - m > 0$ and $n_{\r} > 5$; we have that
    \begin{align}
        \nonumber n_{\r}^4 2^{n_{\r} - m} &> 4n_{\r} 2^{n_{\r}-m}(1 + \frac{1}{3}n_{\r}^2)\\
        \nonumber &> 4n_{\r}(2^{n_{\r}-m} + \frac{1}{3} n_{\r}^2 \sqrt{2}^{n_{\r} - m})\\
        &\geq \ln(2)(n_{\r}+m)2(2^{n_{\r}-m} + \frac{1}{3} n_{\r}^2 \sqrt{2}^{n_{\r} - m}) + 1\ , 
    \end{align} which ensures $2^{n_{\r}+m}p^* < 1$ and completes the proof.}
\end{proof}

\begin{lemma}\label{lem:product SC}
  Suppose that for every $i\in \cal I$ there are two Hermitian operators $C_i, S_i$ acting on a Hilbert space $\A_i$ such that $\pm C_i \leq S_i$, then
  \begin{align}
    \pm \bigotimes_{i\in\cal I} C_i 
    \leq 
    \bigotimes_{i\in \cal I} S_i\ .
  \end{align}
\end{lemma}

\begin{proof}
  For any assignment $\xi_i =\pm$ for all $i\in\cal I$ we have $\bigotimes_i (S_i +\xi_i C_i) \geq 0$. Therefore, if we average this product over all configurations $\{\xi_i \}$ such that $\prod_i \xi_i =1$ we obtain the positive operator
  \begin{align}
    0 \leq \displaystyle \mathop{\mathbb{E}}_{\{\xi_i \}} 
    \bigotimes_i \left(S_i +\xi_i C_i \right)
    = \bigotimes_i S_i +\bigotimes_i C_i\ .
  \end{align}
  Similarly, if we average the product over all configurations $\{\xi_i \}$ such that $\prod_i \xi_i =-1$ then
  \begin{align}
    0 \leq \displaystyle \mathop{\mathbb{E}}_{\{\xi_i \}} 
    \bigotimes_i \left(S_i +\xi_i C_i \right)
    = \bigotimes_i S_i -\bigotimes_i C_i
  \end{align}
  is positive too.
\end{proof}

\section{Proof of \Cref{thm:est-prot}} \label{app:proof-thm-est}
\noindent \textbf{\Cref{thm:est-prot}.} The protocols of \Cref{sec:estimation} generate a {final output} $\rho_{\K\E|\t, \mathbf z}$ satisfying the following security condition
  \begin{align}\label{th:est bound2}
    \sum_{\ \t \in \{0,1\}^n}\!\!
    \sum_{\ \z \in \{0,1\}^{n_{\e}}} \!\!\! \P(\t, \z) 
    \left\| \rho_{\K\E|\t, \z} -
    u_{\K} 
    \rho_{\E|\t, \z} \right\|_1
    \leq  
    \epsilon\ .
  \end{align}
\begin{proof}
The random variables $\t \in \{\mathtt{estimation}, \mathtt{rawbit}\}^n$ are independent and identically distributed according to $(p_\e, p_\r)$.
If in round $l$ we have $t_l = \textsf{estimation}$ then the systems $\A_l$ and $\B_l$ are included in $\A_\e = \bigotimes_{j=1}^{n_\e} \A_j$ and $\B_\e = \bigotimes_{j=1}^{n_\e} \B_j$, and used for estimation.
If $t_l = \mathrm{rawbit}$ then the systems $\A_l$ and $\B_l$ are included in $\A_\r = \bigotimes_{i=1}^{n_\r} \A_i$ and $\B_\r = \bigotimes_{i=1}^{n_\r} \B_i$, and used for generating the {extractor input}.
Without loss of generality we can assume that $\t$ is initially generated before any measurement, and right after, we can re-order the rounds and write the global state as $\rho_{\A_{\r}\B_{\r}\A_{\e}\B_{\e}\E|\t}$.

In each estimation round $j\in \{1, \ldots, n_\e\}$ the pair $\A_j \otimes B_j$ is measured with
\begin{align}
  Q_{z_j} = \sum_{a,b,x,y} 
  \frac 1 4 A_j(a|x) B_j(b|y)\, 
  \delta^{z_j}_{a+b+xy \bmod 2}\ ,
\end{align}
{which has} outcomes $z_j=0,1$. This produces the estimation data $\z = (z_1, \ldots, z_{n_\e})$ distributed according to
\begin{align}
  \P(\mathbf z|\t) 
  =
  \tr\!\!\left[ \rho_{\A_{\e}\B_{\e}|\t} \prod_{j=1}^{n_{\e}} Q_{z_j} \right]\ .
\end{align}
The global state conditioned on a particular value of the estimation data $\z$ is 
\begin{align}\label{rho ArBr}
  \rho_{\A_{\r}\B_{\r} \E|\t,\z}
  =
  \frac 1 {\P(\z|\t) }
  \tr_{\A_{\e}\B_{\e}} \!\!\left[ 
  \rho_{\A_{\r}\B_{\r}\A_{\e}\B_{\e} \E|\t} \!\left(
  \prod_{j=1}^{n_{\e}} Q_{z_j}\! \right)\! \right]\ .
\end{align}
We start by proving the case when using the XOR seedless extractor, as described in \Cref{subsec:xor_prot}\@.
By using \Cref{thm:xor} and the function which defines the output length of the XOR extractor in our spot checking protocol, Equation~\eqref{m(t,z)-xor}, the left-hand side of \Cref{th:est bound2} can be upper bounded by
\begin{align}
  \nonumber
  & \sum_{\t} \sum_{\z} \P(\t, \z) 
  \left\| \rho_{\K\E|\t, \z} -
  u_{\K|\t, \z}
  \rho_{\E|\t, \z} \right\|_1
  \\ \label{step1}\leq &
  \sum_{\t} \sum_{\z} \P(\t, \z)\,  
  m(\t, \z)
  \tr\!\! \left[ \rho_{\A_{\r}\B_{\r} |\t,\z} \prod_{i=1}^{n_{\r}} {S}_{i} \right]\ ,
  \end{align} since, in the case $ m(\t, \z) =0$, the term inside the trace norm is 0 and in the case $m(\t, \z) = 1$, the error can be bounded by \Cref{thm:xor}. Now, using the global state conditioned on a particular value of the estimation data $\z$~\eqref{rho ArBr} and the facts that $\P(\t, \z) = \P(\t)\P(\z|\t) =  p_{\e}^{n_{\e}}\, p_{\r}^{n_{\r}} \P(\z|\t)$ and
    \begin{align} 
        m(\t, \z) &= 
        \begin{cases}
            1 \quad \mathrm{if} \quad  
        \max\limits_{\substack{s \in (2,2\sqrt{2}) \\ \alpha_0, \alpha_1, \beta \in \mathbb{R}}}
        \sum_{j=1}^{n_\e}\alpha_{z_j} + (\beta - 1) n_\r - 2 \log_2(1/\epsilon) \geq 0 \\
            0 \quad \mathrm{otherwise}
        \end{cases},
      \end{align} is upper bounded by the exponential $\sqrt{2}^{\sum_{j=1}^{n_{\e}}\alpha_{z_j} + (\beta-1) n_{\r} - 2 \log_2(1/\epsilon)}$, we bound \Cref{step1}
\begin{align} 
  \nonumber\leq &
  \sum_{\t} \sum_{\z} p_{\e}^{n_{\e}}\, p_{\r}^{n_{\r}}  
  \sqrt{2}^{\sum_{j=1}^{n_{\e}}\alpha_{z_j} + (\beta-1) n_{\r}} \epsilon 
  \tr\!\! \left[ \rho_{\A_{\r}\B_{\r}\A_{\e}\B_{\e} |\t} \prod_{i=1}^{n_{\r}} {S}_{i} \prod_{j=1}^{n_{\e}} Q_{z_j}\!\right]
  \\ \nonumber = &
  \ \epsilon \sum_{\t} 
  \tr\!\!\left[ 
  \rho_{\A_{\r}\B_{\r}\A_{\e}\B_{\e}| \t}
  \prod_{i=1}^{n_{\r}}\! \left(p_{\r} \sqrt{2}^{\beta-1}  {S}_{i} \right) 
  \prod_{j=1}^{n_{\e}}\! \left(p_{\e}\!\left[\sqrt 2^{\alpha_0} Q_{0_j} +\sqrt 2^{\alpha_1} Q_{1_j}\right]\right)\right]
  \\ \label{factors-xor} = &
  \ \epsilon
  \tr\!\!\left[ \rho_{\A_\mathsf{g} \B_\mathsf{g}} \prod_{l=1}^n 
  \left( p_{\r} \sqrt 2^{\beta-1} {S_l}
  +p_{\e} \left[\sqrt 2^{\alpha_0} Q_{0_l} +\sqrt 2^{\alpha_1} Q_{1_l}\right] \right) \right]\ ,
\end{align}
where we denote the global Hilbert spaces of Alice and Bob by $\A_\mathsf{g} = \A_\r \otimes \A_\e$ and $\B_\mathsf{g} = \B_\r \otimes \B_\e$.
Finally, using the following identities
\begin{align} \label{id1}
  S &= \mu_s \unity -4 \nu_s \left(Q_0-Q_1 \right),
  \\ \label{id2}
  \unity &= Q_0+Q_1\ ,
\end{align}
we can write each of the factors in \Cref{factors-xor} as
\begin{align}
  \nonumber
  &\ p_{\r} \sqrt 2^{\beta-1} S 
  +p_{\e} \left[\sqrt 2^{\alpha_0} Q_0 +\sqrt 2^{\alpha_1} Q_1\right]
  \\ \nonumber = &\
  p_{\r} \sqrt 2^{\beta-1} \left[\mu_s(Q_0+Q_1) 
  -4 \nu_s \left(Q_0-Q_1 \right)\right] 
  +p_{\e} \left[\sqrt 2^{\alpha_0} Q_0 +\sqrt 2^{\alpha_1} Q_1\right]
  \\ \nonumber = &
  \left[p_{\r} \sqrt 2^{\beta-1} (\mu_s-4 \nu_s)
  +p_{\e} \sqrt 2^{\alpha_0} \right] Q_0
  +\left[p_{\r} \sqrt 2^{\beta-1} (\mu_s+4 \nu_s)
  +p_{\e} \sqrt 2^{\alpha_1} \right] Q_1
  \\ = \label{b10} &\ Q_0 +Q_1
  = \unity\ ,
\end{align}
where the penultimate equality follows form imposing conditions
\begin{align}
    p_{\r} \sqrt 2^{\beta-1} \left(\mu_s-4 \nu_s \right) 
    +p_{\e} \sqrt 2^{\alpha_0}
    = 1\ ,
    \\
    p_{\r} \sqrt 2^{\beta-1} \left(\mu_s +4 \nu_s \right) 
    +p_{\e} \sqrt 2^{\alpha_1}
    = 1\ ,
\end{align} 
expressed in \Cref{main conditon 1 xor} and \Cref{main conditon 2 xor}\@.
Substituting \Cref{b10} back in \Cref{factors-xor} gives us the bound~\eqref{th:est bound2} and completes the proof.

Similarly, we make the same proof in the $m$-bit extractor function case, as described in \Cref{subsec:balanced_prot}. In this case, we introduce an indicator function 
\begin{align} \label{fun:ind}
    \mathrm{I}(m(\t, \z)) = \begin{cases}
        1 \quad \text{if} \quad m(\t, \z) > 0 \\
        0 \quad \text{otherwise}\ ,
    \end{cases}
\end{align} 
to encode the case when no {output} is produced and the error is $0$. 
By using \Cref{thm-existence} and substituting the global state conditioned on the estimation data~\eqref{rho ArBr}, the output length~\eqref{m(t,z)} and the indicator function~\eqref{fun:ind}, we can write the left-hand side of \Cref{th:est bound2} as follows
\begin{align}
  \nonumber
  & \sum_{\t} \sum_{\z} \P(\t, \z) 
  \left\| \rho_{\K\E|\t, \z} -
  u_{\K|\t, \z}
  \rho_{\E|\t, \z} \right\|_1
  \\ \nonumber \leq &
  \sum_{\t} \sum_{\z} \P(\t, \z)\, \mathrm{I}(m(\t, \z)) 
  \sqrt{2}^{m(\t, \z) -n_{\r} + 4\log_2(n_{\r})}
  \tr\!\! \left[ \rho_{\A_{\r}\B_{\r} |\t,\z} \prod_{i=1}^{n_{\r}} \left(\unity +{ S}_{i}\right) \right]
  \\ \nonumber \leq &
  \sum_{\t} \sum_{\z} \P(\t, \z)\, 
  \sqrt{2}^{m(\t, \z) -n_{\r} + 4\log_2(n_{\r})} 
  \tr\!\! \left[ \rho_{\A_{\r}\B_{\r} |\t,\z} \prod_{i=1}^{n_{\r}} \left(\unity +{ S}_{i}\right) \right]
  \\ \nonumber = &
  \sum_{\t} \sum_{\z} p_{\e}^{n_{\e}}\, p_{\r}^{n_{\r}}  
  \sqrt{2}^{\sum_{j=1}^{n_{\e}}\alpha_{z_j} + (\beta-1) n_{\r}} \epsilon 
  \tr\!\! \left[ \rho_{\A_{\r}\B_{\r}\A_{\e}\B_{\e} |\t} \prod_{i=1}^{n_{\r}} \left(\unity +{S}_{i}\right) \prod_{j=1}^{n_{\e}} Q_{z_j}\!\right]
  \\ \nonumber = &
  \ \epsilon \sum_{\t} 
  \tr\!\!\left[ 
  \rho_{\A_{\r}\B_{\r}\A_{\e}\B_{\e}| \t}
  \prod_{i=1}^{n_{\r}}\! \left(p_{\r} \sqrt{2}^{\beta-1}\!  \left[\unity +{ S}_{i}\right] \right) 
  \prod_{j=1}^{n_{\e}}\! \left(p_{\e}\!\left[\sqrt 2^{\alpha_0} Q_{0_j} +\sqrt 2^{\alpha_1} Q_{1_j}\right]\right)\right]
  \\ \label{factors} = &
  \ \epsilon
  \tr\!\!\left[ \rho_{\A_\mathsf{g} \B_\mathsf{g}} \prod_{l=1}^n 
  \left( p_{\r} \sqrt 2^{\beta-1} \left[\unity +{S}_l\right] 
  +p_{\e} \left[\sqrt 2^{\alpha_0} Q_{0_l} +\sqrt 2^{\alpha_1} Q_{1_l}\right] \right) \right]\ ,
\end{align}
where we again denote the global Hilbert spaces of Alice and Bob by $\A_\mathsf{g} = \A_\r \otimes \A_\e$ and $\B_\mathsf{g} = \B_\r \otimes \B_\e$.
Using the identities for ${S}$ and $\unity$ from \Cref{id1} and \Cref{id2},
we can write each of the factors in \Cref{factors} as
\begin{align}
  \nonumber
  &\ p_{\r} \sqrt 2^{\beta-1} \left[\unity +{ S}\right] 
  +p_{\e} \left[\sqrt 2^{\alpha_0} Q_0 +\sqrt 2^{\alpha_1} Q_1\right]
  \\ \nonumber = &\
  p_{\r} \sqrt 2^{\beta-1} \left[(1+\mu_s)(Q_0+Q_1) 
  -4 \nu_s \left(Q_0-Q_1 \right)\right] 
  +p_{\e} \left[\sqrt 2^{\alpha_0} Q_0 +\sqrt 2^{\alpha_1} Q_1\right]
  \\ \nonumber = &
  \left[p_{\r} \sqrt 2^{\beta-1} (1+\mu_s-4 \nu_s)
  +p_{\e} \sqrt 2^{\alpha_0} \right] Q_0
  +\left[p_{\r} \sqrt 2^{\beta-1} (1+\mu_s+4 \nu_s)
  +p_{\e} \sqrt 2^{\alpha_1} \right] Q_1
  \\ = &\ Q_0 +Q_1
  = \unity\ ,
\end{align}
where the penultimate equality follows from the conditions expressed in \Cref{main conditon 1} and \Cref{main conditon 2}\@.
Substituting this back into \Cref{factors} gives us the bound \Cref{th:est bound2} and completes the proof.
\end{proof}

\end{document}